\documentstyle[preprint,tighten,pra,aps,latexsym,amssymb,floats,epsfig]{revtex}

\def\hs{\qquad}               
\def\nn{\nonumber}            
\def\beq{\begin{eqnarray}}    
\def\eeq{\end{eqnarray}}      
\def\ap{\left.}               
\def\at{\left(}               
\def\aq{\left[}               
\def\cp{\right.}              
\def\ct{\right)}              
\def\cq{\right]}              
\def\ii{\infty}                                    
\def\X{\times\,}                                   
\def\dir{/\kern-.7em D\,}                          

\def\si{\sigma}

\def\Si{\Sigma}
\def\Om{\Omega}

\newcommand{\rif}[1]{(\ref{#1})}
\newcommand{\eq}{\begin{equation}}
\newcommand{\feq}{\end{equation}}
\newcommand{\eqn}{\begin{eqnarray}}
\newcommand{\feqn}{\end{eqnarray}}
\newcommand{\arr}{\begin{eqnarray*}}
\newcommand{\farr}{\end{eqnarray*}}
\newcommand{\lp}{\left(}
\newcommand{\rp}{\right)}

\newcommand{\dM}{{\partial\cal M}}
\newcommand{\M}{{\cal M}}

\newcommand{\dd}{{\rm d}}
\newcommand{\ee}{{\rm e}}

\newcommand{\pdd}[3]{\left.\frac{\partial #1}{\partial #2}\right|_{#3}}
\newcommand{\HP}{Hawking-Page }

\begin{document}

\tightenlines
\draft
\preprint{UTF 434}

\title{Thermodynamics of Kerr-Newman-AdS Black Holes and Conformal
Field Theories}

\author{Marco M.~Caldarelli\footnote{email: caldarel@science.unitn.it},
Guido Cognola\footnote{email: cognola@science.unitn.it}
and Dietmar Klemm\footnote{email: klemm@science.unitn.it}\\ \vspace*{0.5cm}}

\address{Universit\`a  degli Studi di Trento,\\
Dipartimento di Fisica,\\
Via Sommarive 14\\
38050 Povo (TN)\\
Italia\\
\vspace*{0.5cm}      
and\\ Istituto Nazionale di Fisica Nucleare,\\
Gruppo Collegato di Trento,\\ Italia}

\maketitle
\begin{abstract}
We study the thermodynamics of four-dimensional Kerr-Newman-AdS black holes
both in the canonical and the grand-canonical ensemble. The stability
conditions are investigated, and the complete phase diagrams are
obtained, which include the Hawking-Page phase transition in the
grand-canonical ensemble. In the canonical case, one has a first order
transition between small and large black holes, which disappears for
sufficiently large electric charge or angular momentum. This disappearance
corresponds to a critical point in the phase diagram.
Via the AdS/CFT conjecture, the obtained phase structure is also
relevant for the corresponding conformal field theory living in a
rotating Einstein universe, in the presence of a global background
$U(1)$ current. An interesting limit arises when the black holes
preserve some supersymmetry. These BPS black holes correspond to highly
degenerate zero temperature states in the dual CFT, which lives in
an Einstein universe rotating with the speed of light.
\end{abstract}


\maketitle

\section{Introduction}
The conjectured equivalence of string theory on anti-de~Sitter (AdS) spaces
(times some compact manifold) and certain superconformal gauge theories
living on the boundary of AdS
\cite{Maldacena:1997re,Witten:1998qj,Gubser:1998bc,Aharony:1999ti} has lead to
an increasing interest in asymptotically anti-de~Sitter black holes.
This interest is mainly based on the fact that the classical supergravity
black hole solution can furnish important informations on the dual gauge theory
in the large $N$ limit, $N$ denoting the rank of the gauge group.
The standard example of this is the well-known Schwarz\-schild-AdS black hole,
whose thermodynamics was studied 17 years ago by Hawking and Page
\cite{Hawking:1983dh}, who discovered a phase transition from thermal AdS space
to a black hole phase, as the temperature increases. This means that at
a certain temperature thermal radiation in AdS space becomes unstable,
and eventually collapses to form a black hole. The Hawking-Page phase
transition was then reconsidered by Witten \cite{Witten:1998zw}
in the spirit of
the AdS/CFT correspondence. He observed that it can be interpreted as
a transition from a low-temperature confining to a high temperature
deconfining phase in the dual field theory.\\
Up to now, the study of black holes in the context of the AdS/CFT
correspondence has been extended in various directions,
e.~g.~to black holes with non-spherical event horizon topologies 
\cite{Birmingham:1998nr,Caldarelli:1999ar,Emparan:1999gf} or to electrically
charged Reissner-Nordstr\"om-AdS
(RNAdS) solutions \cite{Chamblin:1999tk,Chamblin:1999hg},
following previous work
on this subject performed
in \cite{Louko:1996dw,Peca:1998cs,Mitra:1998tv,Mitra:1999ge}.
Thereby, some interesting
observations have been made, like the striking resemblance of the RNAdS
phase structure to that of the van der Waals-Maxwell liquid-gas system
\cite{Chamblin:1999tk,Chamblin:1999hg}\footnote{A similar phase
structure has recently
been found by considering stringy corrections to AdS black holes
\cite{Caldarelli:1999ar}.}, or the possible appearance of so-called
``precursor" states \cite{Polchinski:1999yd} in the CFT
dual of hyperbolic black
holes \cite{Emparan:1999gf}. Furthermore, the thermodynamics of
R-charged black holes in four, five and seven dimensions has
been studied in detail in \cite{Cvetic:1999ne}.
Another interesting extension is the inclusion
of rotation, i.~e.~the consideration of Kerr-AdS black holes
\cite{Hawking:1998kw,Hawking:1999ab}\footnote{For a discussion
of Kerr-NUT-AdS and
Kerr-bolt-AdS spacetimes cf.~\cite{Mann:1999bt}.}.
In this case, the dual CFT lives in a rotating Einstein universe
\cite{Hawking:1998kw}. The authors of \cite{Hawking:1998kw}
studied the limit of
critical angular velocity, at which this Einstein universe rotates with
the speed of light, from both the CFT and the bulk side. Recently, the
AdS/CFT correspondence has been probed in \cite{Berman:1999mh} by comparing 
in more detail the
thermodynamics of rotating black holes in five-dimensional anti-de~Sitter
space with that of ${\cal N}=4$ super
Yang-Mills theory on a rotating Einstein universe in four dimensions.\\
In the present paper, we shall consider the charged rotating
case, i.~e.~Kerr-Newman-AdS (KNAdS) solutions. (Note that the appearance of
second-order phase transitions for Kerr-Newman black holes in de~Sitter
space, i.~e.~for positive cosmological constant, was first shown by
Davies \cite{Davies:1989ey}).\\
The remainder of this article
is organized as follows:\\
In section \ref{KerrNAdS} we review the four-dimensional KNAdS black hole,
give the conserved quantities like mass and angular momentum, and recall
the conditions under which some amount of supersymmetry is preserved.
In \ref{Eucl} we calculate the Euclidean action, using the method of
counterterms introduced in
\cite{Balasubramanian:1999re,Emparan:1999pm,Kraus:1999di,Mann:1999pc}.
In section \ref{thermo} the black hole thermodynamics is considered in both
the canonical and grand canonical ensemble. The resulting phase
structure is elaborated, and the
thermodynamical stability of the solutions is studied. Finally, in
\ref{concl} our results are summarized and discussed in the context
of the conformal field theory in the rotating Einstein universe.

\section{Kerr-Newman-AdS Black Holes}
\label{KerrNAdS}

Here we consider the usual charged rotating black hole in AdS space.
Its horizon is
homeomorphic to a sphere, and its metric, which is axisymmetric, reads in
Boyer-Lindquist-type coordinates \cite{Carter:1968ab,Plebanski:1976ab}
\eq
ds^2 = -{\Delta_r\over\rho^2}
\left[dt-\frac{a\sin^2\theta}{\Xi}\ d\phi\right]^2
+{\rho^2\over\Delta_r}\ dr^2+{\rho^2\over\Delta_\theta}\ d\theta^2
+{\Delta_\theta\sin^2\theta\over\rho^2}
\left[a\ dt-\frac{r^2+a^2}{\Xi}\ d\phi\right]^2,
\label{KNAdS}
\feq
where
\eq 
\rho^2=r^2+a^2\cos^2\theta, \qquad \Xi=1-{a^2\over l^2},
\feq
\eq
\Delta_r=(r^2+a^2)\lp 1+{r^2\over l^2}\rp-2mr+z^2 , \qquad 
\Delta_\theta=1-{a^2\over l^2}\cos^2\theta\:.
\feq
Here $a$ denotes the rotational parameter and $z$ is defined
by $z^2=q_e^2+q_m^2$,
$q_e$ and $q_m$ being electric and magnetic charge parameters respectively. 

The metric (\ref{KNAdS}) solves the Einstein-Maxwell field equations with an
electromagnetic vector potential and an associated field strength tensor
respectively given by
\eq
A = -{q_er\over\rho\sqrt{\Delta_r}}\ {\rm e}^0
-{q_m\cos\theta\over\rho\sqrt{\Delta_\theta}\sin\theta}\ {\rm e}^3,
\feq
and
\begin{eqnarray}
F&=&-\frac 1{\rho^4}\left[ q_e(r^2-a^2\cos^2\theta)+2q_mra\cos\theta\right]
\ \ee^0\wedge\ee^1\nonumber\\
&&\hs+\frac 1{\rho^4}\left[q_m(r^2-a^2\cos^2\theta)-2q_era\cos\theta\right]
\ \ee^2\wedge\ee^3,\label{fieldKNAdS}
\end{eqnarray}
where we have defined the vierbein field
\eq
  \ee^0={\sqrt{\Delta_r}\over\rho} (\dd
  t-\frac{a\sin^2\theta}{\Xi}\ \dd\phi),\qquad
  \ee^1={\rho\over\sqrt{\Delta_r}}\ \dd r,
\feq
\eq
  \ee^2={\rho\over\sqrt{\Delta_\theta}}\ \dd\theta,\qquad
  \ee^3={\sqrt{\Delta_\theta}\sin\theta\over\rho} (a\ \dd t-\frac{r^2+a^2}\Xi
  \ \dd\phi).
\feq
Let us define the critical mass parameter $m_{extr}$,
\eqn
m_{extr}(a,z)&=&{l\over3\sqrt6}\lp\sqrt{\lp1+{a^2\over l^2}\rp^2+{12\over
    l^2}(a^2+z^2)}+{2a^2\over l^2}+2\rp\nonumber\\
&&\hs\times\lp\sqrt{\lp1+{a^2\over l^2}\rp^2+{12\over
    l^2}(a^2+z^2)}-{a^2\over l^2}-1\rp^{1\over 2}.
\label{extrKNAdS}\feqn
A study of the positive zeroes of the function $\Delta_r$ shows
that the line element \rif{KNAdS} describes a
naked singularity for $m<m_{extr}$\footnote{Note that the solution with
$m=z=0$ describes AdS space seen by a rotating observer.}
and a
black hole with an outer event horizon and an inner Cauchy horizon for
$m>m_{extr}$. Finally, for $m=m_{extr}$, the lapse function has a double root
and \rif{KNAdS} represents an extremal black hole. 
One should also note that the solution is valid only for $a^2<l^2$; it
becomes singular in the limit $a^2=l^2$. In this critical limit, which
has been studied extensively in Ref.~\cite{Hawking:1998kw} (cf.~also
\cite{Berman:1999mh}), the three dimensional Einstein universe at infinity
rotates with the speed of light.
Here we assume $m$ to be larger than $m_{extr}$ and $a<l$, so  
the metric (\ref{KNAdS}) represents an AdS black hole,
with an event horizon at $r=r_+$, where $r_+$ is the largest solution of
$\Delta_r=0$. For the horizon area one gets
\eq
{\cal A} = \frac{4\pi(r_+^2+a^2)}{\Xi}. \label{horarea}
\feq
Analytical continuation of the Lorentzian metric by
$t\to i\tau$ and $a\to ia$ yields the Euclidean section, whose regularity
at $r=r_+$ requires that we must identify $\tau \sim \tau + \beta$
and $\phi \sim \phi + i\beta\Omega_H$, where the inverse Hawking temperature
$\beta$ is given by
\eq
\beta=\frac{4\pi(r_+^2+a^2)}{r_+\left(1+\frac{a^2}{l^2}+3\frac{r_+^2}{l^2}
-\frac{a^2+z^2}{r_+^2}\right)}, \label{beta}
\feq
and
\eq
\Om_H=\frac{a\Xi}{r_+^2+a^2}
\feq
represents the angular velocity of the event horizon.
$\Om_H$ can also be
obtained by using the fact that the surface $r=r_+$ is a bolt of the
co-rotating
Killing vector $\partial_\tau+i\Om_H\partial_\phi$ in the Euclidean
section.
One can also write the metric \rif{KNAdS} in the canonical form
\eq
ds^2 = -N^2dt^2 + \frac{\rho^2}{\Delta_r}dr^2 + \frac{\rho^2}{\Delta_{\theta}}
       d\theta^2 + \frac{\Sigma^2\sin^2\theta}{\rho^2\Xi^2}(d\phi - \omega
       dt)^2, \label{canform}
\feq
where
\eq
\Sigma^2 = (r^2+a^2)^2\Delta_{\theta} - a^2\Delta_r\sin^2\theta,
\feq
and the lapse function $N$ and the angular velocity $\omega$
are defined by
\eqn
N^2 &=& \frac{\rho^2\Delta_r\Delta_{\theta}}{\Sigma^2}, \\
\omega &=& \frac{a\Xi}{\Sigma^2}\left[\Delta_{\theta}(r^2+a^2) -
           \Delta_r\right] \label{om}
\feqn
respectively. Note that $\omega = \Omega_H$ on the horizon, whereas
$\omega = -a/l^2 \equiv \Omega_{\infty}$ for $r\to\infty$.
The fact that the angular velocity
does not vanish at infinity is a salient feature of rotating black holes
in AdS space, different from the asymptotically flat case, where
$\Omega_{\infty} = 0$. The angular velocity $\Omega$
entering the thermodynamics is then given by the {\it difference}
\eq
\Omega = \Omega_H - \Omega_{\infty}, \label{differ}
\feq
analogously to the case of an electric potential. For asymptotically flat
spacetimes, \rif{differ} reduces to $\Omega = \Omega_H$, but for the
KNAdS black hole we get
\eq
\Om=\Om_H+\frac{a}{l^2}=\frac{a(1+r_+^2/l^2)}{r_+^2+a^2}\:. \label{OmEinst}
\feq
This is exactly the angular velocity of the rotating Einstein universe
at infinity \cite{Hawking:1998kw}. 
To see this, one first applies
an implicit coordinate transformation,
which takes the standard AdS metric to the $m=z=0$ KNAdS form
\cite{Hawking:1998kw}. \rif{OmEinst} can then be read off from the
transformation rule of the coordinate $\phi$.\\
The fact that the angular velocity relevant to Kerr-Newman-AdS black hole
thermodynamics turns out to be the one of the rotating Einstein universe
at the AdS boundary, agrees nicely with the AdS/CFT correspondence:
If the KNAdS black hole in
the bulk is described by a conformal field theory living on the boundary,
then the relevant angular velocity entering the thermodynamics should be
that of the rotating Einstein universe at infinity.\\
Moreover, when $\Omega<1/l$, a timelike killing vector can be
globally defined outside the event horizon \cite{Hawking:1998kw}, in
contrast to the asymptotically flat case.
Then the rotating black hole
can be in equilibrium with rotating thermal radiation all the way out to 
infinity, and a consistent thermodynamics can be defined. When this condition
is fullfilled, the rotating Einstein universe spins slower than light, and
an associated state can be defined on the boundary CFT. This is closely
related with the fact that for $\Omega<1/l$ there are no superradiant modes
\cite{Hawking:1999ab}.\\
The mass $M$ and the angular momentum $J$ can be defined by 
means of Komar integrals, using the Killing vectors 
$\partial_t/\Xi$ and $\partial_\phi$\footnote{We have normalized the Killing
vectors so that the corresponding conserved quantities generate the
algebra $so(3,2)$, like it was done in Ref.~\cite{Kostelecky:1996ei}.}.
Taking AdS space as reference
background, one gets
\eq
M=\frac{m}{\Xi^2},\qquad
J=\frac{am}{\Xi^2}. \label{MJ}
\feq
The charges are obtained by computing the flux of the electromagnetic 
field tensor at infinity, yielding
\eq
Q_e=\frac{q_e}\Xi,\qquad Q_m=\frac{q_m}\Xi. \label{Q}
\feq
The electric potential $\Phi$, measured at infinity with respect to
the horizon, is defined by
\eq
\Phi = A^{e}_a\chi^a|_{r\to\infty} -  A^{e}_a\chi^a|_{r=r_+} =
       \frac{q_er_+}{r_+^2+a^2}, \label{Phi1}
\feq
where $\chi = \partial_t + \Omega_H\partial_{\phi}$ is the null
generator of the horizon and $A^{e}_a$ denotes the electric part
of the vector potential.\\
Finally, let us recall the conditions under which some amount of
supersymmetry is preserved by the Kerr-Newman-AdS black holes.
These objects were considered in \cite{Kostelecky:1996ei,Caldarelli:1998hg}
in the context
of gauged $N=2$, $D=4$ supergravity. There it was found that for
\eq
m^2 = la\left(1+\frac al\right)^4, \qquad q_e^2 = la\left(1+\frac al\right)^2,
      \qquad q_m = 0
\label{susy}
\feq
the KNAdS black holes preserve half of the supersymmetries.
In the BPS limit we have
\eq
M = \frac{\sqrt{la}}{\left(1-\frac al\right)^2}, \qquad Q_e =
\frac{\sqrt{la}}{1-\frac al}, \qquad Q_m = 0, \qquad J =
\frac{\sqrt{la}a}{\left(1-\frac al\right)^2},
\feq
so the Bogomol'nyi bound is \cite{Kostelecky:1996ei}
\eq
M = Q_e + \frac Jl.
\feq
Furthermore, in the supersymmetric case, one has the relations
\eq
r_+^2 = al, \qquad \Om = \frac 1l. 
\feq
The latter relation means that in the BPS limit, the Einstein universe
at infinity, in which the dual conformal field theory lives, rotates
effectively with the speed of light. An analogous situation was found in
Ref.~\cite{Hawking:1998kw} for the supersymmetric rotating BTZ black hole.

\section{Euclidean Action}
\label{Eucl}

We turn now to the calculation of the Euclidean action, which will
then yield the thermodynamic potentials relevant to the various ensembles
\cite{Gibbons:1977ue}.
In the action computation one usually encounters infrared divergences,
which are regularized by subtracting a suitably chosen background.
Such a procedure, however, in general is not unique; in some cases
the choice of reference background is ambiguous, e.~g.~for hyperbolic
AdS black holes \cite{Vanzo:1997gw,Emparan:1999pm}.
Recently, in order
to regularize divergent integrals like those appearing in the computation
of the Euclidean action,
a different procedure has been proposed
\cite{Balasubramanian:1999re,Emparan:1999pm,Kraus:1999di,Mann:1999pc}.
This technique was inspired by the AdS/CFT correspondence, and
consists in adding suitable counterterms $I_{ct}$ to the action. These 
counterterms are built up with curvature invariants of a boundary
$\partial {\cal M}$ (which is sent to infinity after the integration),
and thus obviously do not alter the bulk equations of motion.
This kind of procedure, which will also be employed in the present paper,
has the advantage to be free of ambiguities,
which, on the contrary, are present in the traditional approach 
in some particular cases, like mentioned above.

To start with, we write the Euclidean Einstein-Maxwell action in the form
\beq
I=I_{bulk}+I_{surf}+I_{ct}&=&
-\frac1{16\pi G}\int_{\cal M}\!d^4x\ \sqrt{g}\left[R-2\Lambda-F^2\right]
\nn\\&&\hs
-\frac1{8\pi G}\int_{\partial{\cal M}}\!d^3x\ \sqrt{h}\ K+I_{ct}\:,
\label{action}
\eeq
$\Lambda=-3/l^2$ being
the 4-dimensional cosmological constant, with
$l$ the radius of AdS space.
The action (\ref{action})
differs from the familiar one by the presence
of the last term, which contains all surface counterterms needed to assure 
convergence of the integrals. In four dimensions it reads \cite{Emparan:1999pm}
\beq
I_{ct}=\int_{\partial{\cal M}}\!d^3x\ \sqrt{h}
\aq
\frac2l+\frac l2{\cal R}
-\frac{l^3}2\at{\cal R}_{ab}{\cal R}^{ab}-\frac38{\cal R}^2\ct
\cq,\label{actionCT}\eeq 
$h_{ab}$ and ${\cal R}_{ab}$ denoting the metric and Ricci curvature of
the (arbitrary) boundary $\partial{\cal M}$.

The variation of the bulk integral with respect to the fields gives
\eqn
\delta I&=&-\frac1{16\pi G}\int_\M\!d^4x\ \sqrt{g}\left[G_{ab}
+\Lambda g_{ab}-\left(2F_a{}^cF_{bc}
-\frac12F^2g_{ab}\right)\right]\delta g^{ab}
\nonumber\\
&&-\frac1{4\pi G}\int_\M\!d^4x\ \sqrt{g}\lp\nabla_aF^{ab}\rp\delta A_b
+\frac1{4\pi G}\int_\dM\!d^3x\ \sqrt{h}\ n_aF^{ab}\delta A_b,
\label{varia}
\feqn
where $n^a$ is the outward pointing unit normal to $\dM$. 
Asking for a stationary point of the action, the bulk integrals
in Eq.~(\ref{varia}) yield the Einstein-Maxwell field equations
\eq
G_{ab}+\Lambda g_{ab}=T_{ab}\,,\qquad \nabla_aF^{ab}=0,
\feq
where we have defined the electromagnetic stress tensor
\eq
T_{ab}=2F^c{}_aF_{cb}-\frac12F^2 g_{ab}\,,
\feq
while the surface integral has to vanish to have a differentiable action
functional and a well defined action principle. This imposes the boundary
condition $\delta A_a=0$ on $\dM$.
This action is thus appropriate to study the ensemble with fixed electric
potential $\Phi_e$ and fixed magnetic charge $Q_m$.\\
To study the canonical ensemble with fixed magnetic {\em and} electric
charge, we have to add another boundary term to impose fixed
$Q_e$ as a boundary condition at infinity \cite{Hawking:1995ap}, restoring
the electromagnetic duality.
The appropriate action in this case is
\eq
\tilde I = I-\frac1{4\pi G}\int_\dM\!d^3x\ \sqrt{h}\ n_aF^{ab}A_b,
\feq
and its variation reads
\eq
\delta\tilde I=({\rm bulk\,\,terms})
-\frac1{4\pi G}\int_\dM\!d^3x\ \sqrt{h}\ A^b\delta\lp n^aF_{ab}\rp.
\feq
The bulk contribution yields again the Einstein-Maxwell equations, while
the vanishing of the surface term, needed to have a differentiable action
functional,  requires $\delta(n^aF_{ab})=0$ at infinity as a boundary
condition.\\

Now the evaluation of the Euclidean action is a straightforward computation.
First of all, on shell the bulk contribution simplifies to 
\eq
I_{bulk}=\frac1{16\pi G}\int\!d^4x\ \sqrt{g}\left[F^2+\frac6{l^2}\right].
\feq
Then one chooses the boundary $\dM=S^1\X S^2$, 
$S^2$ being a 2-sphere with a large radius, 
which has to be sent to infinity after the integration.
Since the metric is stationary, the time integration gives rise to a
simple multiplicative factor $\beta$.
The integration on the other variables requires a little bit of work, 
but anyway it can be performed in a closed manner.
The final results assume the form
\eqn
I=\frac\beta{4Gl^2\Xi}
\left[-r_+^3+\Xi l^2r_+ +\frac{l^2a^2}{r_+}
+\frac{l^2(q_e^2+q_m^2)}{r_+}
-2\frac{l^2(q_e^2-q_m^2)r_+}{r_+^2+a^2}
\right]
\:,\label{aactionGC}
\feqn
\eqn
\tilde I=\frac\beta{4Gl^2\Xi}
\left[-r_+^3+\Xi l^2r_+ +\frac{l^2(a^2+z^2)}{r_+}
+2\frac{l^2z^2r_+}{r_+^2+a^2}
\right]
\:,\label{aactionC}
\feqn
valid for fixed potential and fixed charge respectively. 
In the following, we will consider only the case of vanishing
magnetic charge, $q_m=0$, so we set $q_e = q$ and $Q_e = Q$.

The behaviour of the obtained actions as functions of $r_+$ determines
the thermodynamical properties of the black holes. It will be
discussed in detail in section \ref{thermo}.
Note that both actions are singular
in the extremal case ($m=m_{extr}$) and also for $\Xi=0$. The singularity
for extremal black holes comes from the divergence of the inverse
temperature $\beta$. Since one divides the action by $\beta$ in order
to obtain the thermodynamical potentials, the latter are well-defined
also for $T=0$.\\
In the last section we have already indicated how to calculate
physical conserved quantities by evaluating Komar integrals.
This traditional technique involves the subtraction of a suitably chosen
reference background (in our case AdS space), in order to regularize
infrared-divergent integrals.
In the approach we are considering, instead, one does not use any
reference background, and so it seems quite natural to compute the
conserved charges by the direct use of the finite action (\ref{action}),
applying the method developed by Brown and York \cite{Brown:1993br}.

Let us briefly describe how this formalism works.
We indicate by $u^a$ the unit normal vector of a spacelike 
hyper-surface $^3{\cal S}_t$ at constant $t$ and by $\partial {\cal M}$
the spatial boundary of
the spacetime manifold $\M$, with induced metric $h_{ab}$.
Moreover, $\Si=S^2$ is the spacelike intersection $^3{\cal S}_t\cap
\partial {\cal M}$ embedded in $\partial {\cal M}$ with induced
metric $\si_{ab}$.

One starts by deriving the local surface energy-momentum stress tensor
\beq
\tau^{ab}=\frac2{\sqrt{-h}}\frac{\delta I}{\delta h_{ab}}\:,
\label{tau}
\eeq 
which characterizes the entire system, including contributions 
from gravitation and (if present) matter.
It is related to the stress energy tensor 
$T^{ab}=\frac2{\sqrt{-g}}\frac{\delta I}{\delta g_{ab}}$ 
in a rather complicated way.

Then, for any Killing vector field $\xi^a$ associated with an isometry
of the boundary three-metric, one defines the conserved charge
\footnote{Note that in the conventions of \cite{Brown:1993br} an
additional minus sign appears in the conserved charges (\ref{charges});
this comes from the fact that the definition of the extrinsic curvatures
in \cite{Brown:1993br} differs from ours by a factor of $-1$.}
\beq
Q_{\xi}=\int_{\Si}\:d^2x\:\sqrt{\si}\:u_a\tau^{ab}\xi_b\:. \label{charges}
\eeq
Now, a straightforward application of (\ref{tau}) 
with the action (\ref{action}) gives
\beq 
\tau^{ab}=-\frac1{8\pi G}\aq\at K^{ab}-h^{ab}K\ct
+\frac2l h^{ab}-l\at{\cal R}^{ab}-\frac12 h^{ab}{\cal R}\ct
\cq\:.\label{tauab}
\eeq
The first term on the right-hand side of Eq.~\rif{tauab} results
from the boundary term in the action, while all other terms are due
to the presence of the counterterms we added in order to have 
finite quantities, when the boundary is sent to infinity. 
Employing the AdS/CFT correspondence, the result (\ref{tauab}) 
can also be interpreted as the expectation value of the stress tensor
in the boundary conformal field theory.
In the case of the Kerr-Newman-AdS geometry, we choose $\partial {\cal M}$
to be a three-surface of fixed $r$, and obtain for $\tau^{ab}$
\eqn
8\pi G\tau_{tt} &=& \frac{2m}{rl} + {\cal O}(1/r^2)\:, \nonumber \\
8\pi G\tau_{t\phi} &=& -\frac{2ma}{rl\Xi}\sin^2\theta +
                       {\cal O}(1/r^2)\:, \nonumber \\
8\pi G\tau_{\phi\phi} &=& \frac{m}{rl\Xi^2}\sin^2\theta [l^2 +
                          3a^2\sin^2\theta - a^2] + {\cal O}(1/r^2)\:,
                          \nonumber \\
8\pi G\tau_{\theta\theta} &=& \frac{ml}{r\Delta_{\theta}} +
                              {\cal O}(1/r^2)\:,
\feqn
all other components vanishing.
As expected, the presence of counterterms cures also the divergences of
the charges. In fact we get  
\beq 
M=Q_{\partial_t/\Xi}=\frac{m}{\Xi^2}\:,
\eeq
\beq 
J=Q_{\partial_\phi}=\frac{am}{\Xi^2}\:,
\eeq
in full agreement with the results (\ref{MJ}) obtained with Komar integrals.

\section{Thermodynamics}
\label{thermo}

\subsection{Generalized Smarr Formula}
\label{smarr}

Using the expressions (\ref{MJ}) for mass and angular momentum,
(\ref{horarea}) for the horizon area, and the fact that $\Delta_r=0$
for $r=r_+$, one obtains by simple algebraic manipulations
a generalized Smarr formula for Kerr-Newman-AdS
black holes, which reads
\eq
M^2=\frac{\cal A}{16\pi}+\frac{\pi}{\cal A}(4J^2+Q^4)+\frac{Q^2}2+
\frac{J^2}{l^2}+\frac{\cal A}{8\pi l^2}\left(Q^2+\frac{\cal A}{4\pi}
+\frac{{\cal A}^2}{32\pi^2l^2}\right).
\label{smarr1}\feq
In the limit $l\to\infty$ this formula reduces to the usual Smarr
formula for asymptotically flat Kerr-Newman solutions 
\cite{Smarr,Davies:1977ab},
so one can consider the last two terms on the right-hand side as
AdS corrections.\\
Relation \rif{smarr1} holds for classical black holes, and contains as
usual all the information about the thermodynamic state of the black hole,
taking the black hole entropy to be one quarter of the
horizon area\footnote{In the following we set $G=1$.},
\eq
S=\frac{{\cal A}}{4}. \label{bekhawkentr}
\feq
In the next subsection, we shall check Eq.~(\ref{bekhawkentr})
using the
Euclidean actions (\ref{aactionGC}), (\ref{aactionC}), and standard
thermodynamical relations.\\
We may then regard the parameters $S$, $J$ and $Q$ as a complete set of
energetic extensive parameters for the black hole thermodynamical
fundamental relation $M=M(S,J,Q)$,
\eq
M^2=\frac{S}{4\pi}+\frac{\pi}{4S}(4J^2+Q^4)+\frac{Q^2}2+
\frac{J^2}{l^2}+\frac{S}{2\pi l^2}\left(Q^2+\frac{S}{\pi}
+\frac{S^2}{2\pi^2l^2}\right).
\label{smarr2}
\feq
One can then define the quantities conjugate to $S$, $J$ and $Q$.
These are the temperature
\eq
T=\pdd MS{JQ}=\frac1{8\pi M}\left[1-\frac{\pi^2}{S^2}\lp4J^2+Q^4\rp
+\frac2{l^2}\lp Q^2+\frac{2S}\pi
\rp+\frac{3S^2}{\pi^2l^4}\right]\:,
\label{T}
\feq
the angular velocity
\eq
\Omega=\pdd MJ{SQ}=\frac {\pi J}{MS}\lp1+\frac S{\pi l^2}\rp\:,
\label{Omega}
\feq
and the electric potential
\eq
\Phi=\pdd MQ{SJ}=\frac{\pi Q}{2MS}\lp Q^2+\frac S\pi+\frac{S^2}{\pi^2 l^2}
                 \rp\:.
\label{Phi}
\feq
These expressions reduce to the corresponding asymptotically flat Kerr-Newman
expressions in the
$l\to\infty$ limit \cite{Davies:1977ab}. The obtained quantities satisfy the
first law of thermodynamics
\eq
\dd M=T\dd S+\Omega\dd J +\Phi\dd Q.
\feq
Furthermore, by eliminating $M$ from \rif{T}--\rif{Phi}
using \rif{smarr2}, it is possible to obtain three equations of
state for the KNAdS black holes.\\
One easily verifies that the relations \rif{T}, \rif{Omega} and
\rif{Phi} for temperature, angular velocity and electric potential
respectively,
coincide with equations \rif{beta}, \rif{OmEinst} and \rif{Phi1} found in
section \ref{KerrNAdS}.\\
Another quantity of interest is the thermal capacity $C_{J,Q}$ at constant
angular momentum and charge, which, as we shall see in section \ref{stability},
is relevant in the stability
analysis of the canonical ensemble. It reads
\begin{eqnarray}
\displaystyle C_{J,Q}&\displaystyle =
T\pdd ST{JQ}&=\frac{4\pi^{-1}MTS^3}{4J^2+Q^4-4T^2\frac{S^3}\pi
+\frac{2S^3}{\pi^3l^2}+\frac{3S^4}{\pi^4l^4}} \nonumber\\
&&\displaystyle=\frac{4\pi MTS}{1-4\pi T\lp2M+TS\rp+\frac2{l^2}\lp Q^2+\frac{3S}\pi
\rp+\frac{6S^2}{\pi^2l^4}}\:.
\label{CJQ}\end{eqnarray}
In the above discussion, we have treated the cosmological constant as a fixed
parameter; however, as shown by Henneaux and Teitelboim \cite{Henneaux:1985tv},
it is possible to
induce it from a three form gauge potential coupled to the gravitational
field. The interest of this mechanism resides in that it follows from
Kaluza-Klein reduction of supergravity theories \cite{Duff:1986hr},
and in particular it is
relevant for the compactification of M-theory on $S^7$. We can thus
promote the cosmological constant to a thermodynamic state variable. Its
conjugate variable $\Theta$ reads
\eq
\Theta=\pdd M\Lambda{SJQ}=-\frac1{2M}\left[\frac13J^2+\frac S{6\pi}
\lp Q^2+\frac S\pi\rp-\frac{\Lambda S^3}{18\pi^3}\right].
\feq
Now we can consider ensembles where $\Lambda$ is allowed to fluctuate, and
the first law reads
\eq
\dd M=T\dd S+\Omega\dd J +\Phi\dd Q+\Theta\dd\Lambda.
\feq
Finally, if we regard $M$ as a function of $S$, $J$, $Q^2$ and $\Lambda$, 
it is a homogeneous function of degree $1/2$. Applying Euler's theorem we
obtain
\begin{eqnarray}
\frac12M&=&TS+\Omega J+\frac12\Phi Q-\Theta\Lambda\nonumber\\
&=&TS+\Omega J+\frac12\Phi Q-\frac S{2\pi l^2}\left[\frac{\Phi S}
{\pi Q}+\frac{\Omega J}{1+\frac S{\pi l^2}}\right].
\label{smarr3}\end{eqnarray}
Again, sending $l$ to infinity we recover the usual Smarr law, and the last
term is an AdS correction.

\subsection{Thermodynamic Potentials from the Euclidean Action}
\label{potentials}

We now turn to the definition of the thermodynamic potentials
from the Euclidean actions \rif{aactionC} and \rif{aactionGC},
relevant for the canonical and the grand-canonical ensemble respectively.
Let us first treat the latter case. The Gibbs potential $G(T,\Om,\Phi)$
is defined by
\eq
G(T,\Om,\Phi) = \frac{I}{\beta}, \label{Gibbs}
\feq
with $I$ given by \rif{aactionGC}. Using the expressions \rif{beta} for
the temperature, \rif{OmEinst} for the angular velocity,
and \rif{Phi1} for the electric potential,
we get after some algebra the corresponding extensive quantities
\eq
S = -\pdd GT{\Om\Phi}, \qquad
J = -\pdd G{\Om}{T\Phi}, \qquad
Q = -\pdd G{\Phi}{T\Om},
\feq
which turn out to coincide precisely with the expressions \rif{bekhawkentr},
\rif{MJ} and \rif{Q}. 
One further readily verifies that
\eq
G(T,\Om,\Phi) = M - \Omega J - TS - \Phi Q,
\feq
which means that $G$ is indeed the Legendre transform of the energy
$M(S,J,Q)$ (cf.~\rif{smarr2}) with respect to $S,J$ and $Q$.\\
As for the canonical ensemble, the Helmholtz free energy is
defined by
\eq
F(T,J,Q) = \frac{\tilde{I}}{\beta} + \Omega J, \label{Helmholtz}
\feq
with $\tilde{I}$ given by \rif{aactionC}. The term $\Omega J$ comes 
from the needed Legendre transformation to hold the angular momentum 
fixed. Using the relations \rif{beta}
for the temperature, \rif{MJ} for the angular momentum, and \rif{Q}
for the electric charge, one verifies that the conjugate quantities
\eq
S = -\pdd FT{JQ}, \qquad
\Om = \pdd F{J}{TQ}, \qquad
\Phi = \pdd F{Q}{TJ},
\feq
agree with expressions \rif{bekhawkentr}, \rif{OmEinst} and \rif{Phi1}.
Furthermore one easily shows that
\eq
F(T,J,Q) = M - TS,
\feq
so $F$ is in fact the Legendre transform of $M(S,J,Q)$ with respect to $S$.

\subsection{Stability Analysis}\label{stability}

The stability of a thermodynamical system with respect to small variations  
of the thermodynamic coordinates, is usually studied by analyzing
the behaviour of the entropy $S(M,J,Q)$. The stability requires that the 
entropy hypersurface lies everywhere below its family of tangent
hyperplanes, as to say that the entropy must be a concave function
of the entropic extensive parameters.
This will place some restrictions on physical observables.
For example, the thermal capacity must be positive in any stable system. 
The stability can also be studied by using other thermodynamic potentials,
as for
example the energy or its Legendre transforms, which have to be 
convex functions of their extensive variables, and concave functions
of their intensive variables.
The use of one thermodynamic potential with respect to another 
is of course a matter of convenience, depending on the
ensemble one is dealing with.
Here we are mainly interested in finding the zones where the system 
is locally stable. These are bounded by ``critical'' hypersurfaces,
on which 
\eq
\det\left(\frac{\partial^2 S}{\partial X_i \partial X_j}\right) = 0,
\feq
where $X_i = M,J,Q$.\\
In the canonical ensemble, charges and angular momentum are fixed parameters
and for this reason, the positivity of the thermal capacity $C_{J,Q}$ is
sufficient to assure stability \cite{Cvetic:1999ne}.
This means that the points (hypersurfaces)
where $C_{J,Q}$ vanishes or diverges
represent the critical hypersurfaces we are looking for. 
The equation of such surfaces can easily be obtained by deriving the expression
for $T$, Eq.~(\ref{T}) (where $M$ is given by \rif{smarr2}), with
respect to $S$ and requiring it to vanish
(or to diverge). One can see that the equation
$\partial T/\partial S=0,\ii$ has only one physically acceptable solution,
which reads
\footnote{In the following we set $l=1$.}
\beq 
J^2&=&
-\frac{3{S^2}
        {{( \pi+S ) }^2}
        ( \pi+2S )+
       {\pi^4}{Q^4}
        ( 3\pi+2S )+
       2{\pi^3}{Q^2}S
        ( 2\pi+3S )  ) }{4{\pi^4}
     ( 3\pi+4S )}
\nn\\&&\hs
+\frac{S}{2{\pi^4}(3\pi+4S)}
\aq\phantom{\frac{}{}} 
       {\pi^8}{Q^8}+
         6{\pi^3}{Q^2}S
          {{( \pi+S ) }^4} 
           -2{\pi^7}{Q^6}(2\pi+3S )
\cp\nn\\&&\hs\hs\hs
     +{S^2}{{( \pi+S ) }^3}
      ( 3{\pi^3}+10{\pi^2}S+
      15\pi {S^2}+9{S^3} ) 
\nn\\&&\ap\hs\hs\hs\hs\phantom{\frac{}{}}
        +{\pi^4}{Q^4}
          ( 4{\pi^4}+9{\pi^3}S+
            6{\pi^2}{S^2}+
            6\pi {S^3}+6{S^4} )
\cq^\frac12\:.
\label{J2}
\eeq
In the grand-canonical ensemble the critical surfaces can be determined by 
requiring that the determinant of the whole Hessian of the Gibbs potential
$G(T,\Om,\Phi)$ be vanishing (diverging) \cite{Cvetic:1999ne}.
This yields an equation which can easily be
resolved for the charge parameter $q$. The result is
\beq 
q^2=-{\frac{-{a^4}-2\,{a^2}\,{r_+^2}- 2\,{a^4}\,{r_+^2}-{r_+^4} 
-{a^4}\,{r_+^4}+2\,{r_+^6}+2\,{a^2}\,{r_+^6}
 +3\,{r_+^8}}{-3\,{a^2}+{r_+^2}+{a^2}\,{r_+^2}+{r_+^4}}}.
\label{qe2}
\eeq
These stability conditions will be analyzed in the following subsections.

\subsection{Canonical Ensemble}

In the canonical ensemble, we study the black holes holding the temperature
$T$, the angular momentum $J$ and the charge $Q$ fixed. The associated
thermodynamic potential is the Helmholtz free energy $F(T,J,Q)$
\rif{Helmholtz}. We shall analyze the thermodynamics in this ensemble
in the $(T,J)$-plane, keeping the charge fixed. It immediately follows that
in this plane the extremal black hole solutions are represented by the
ordinate axis $T=0$, and the half-plane $T>0$ corresponds to the
nonextremal solutions.\\
We shall begin by considering the local stability of the black holes. As
explained in Section \ref{stability}, local stability in the canonical
ensemble is equivalent to the positivity of the thermal capacity
\rif{CJQ}. The behaviour of the thermal capacity can be more easily
understood from the state equation $T=T(S,J,Q)$ at fixed $J$ and
$Q$, Eq. \ref{T}. 
\begin{figure}[htb]
\centerline{
\epsfig{file=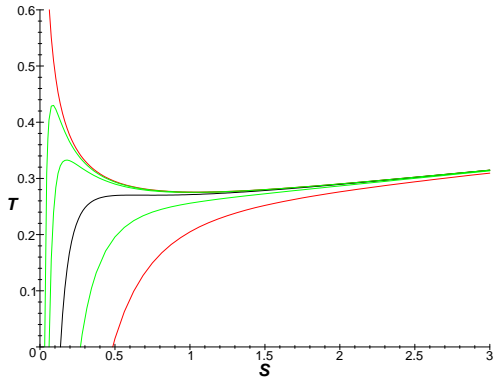,angle=-90,width=0.5\linewidth}}
\caption{{\sf State equation $T=T(S)$.} The plot shows the $T(S)$ curves at
  fixed $Q=0$ for $J=0$ (the upper curve), $0.005$, $0.01$, $0.0236$,
  $0.05$ and $0.1$ (the lower curve). For $J_c\approx0.0236$ the local
  extrema merge in a point of inflection, and for higher $J$ the curve is
  monotonically increasing. For $Q\neq0$, see the text.}
\label{fig:ST}
\end{figure}
In figure \ref{fig:ST} we have reported these curves for $Q=0$ and
different values of $J$.
When $J=0$, the usual Schwarzschild-AdS behaviour
is reproduced; the curve first decrases towards a minimum, corresponding to
the branch of small unstable black holes, then increases along the branch of
large stable black holes. As soon as $J\neq 0$, a branch of stable small
black holes appears, separated from the branch of large black holes by
intermediate unstable black holes. This results in the appearance of two
phases in the canonical ensemble, a small black hole phase and a large
black hole phase, in analogy to what was found in \cite{Chamblin:1999tk}
for charged AdS black holes and in \cite{Caldarelli:1999ar} for
stringy-corrected black holes. While $J$ grows, the local maximum of $T(S)$
decreases, and eventually degenerates for $J_c\approx 0.0236$ into a
point of inflection,
signalling a second order phase transition for the Kerr-AdS black hole. For
$J>J_c$, the curve $T(S)$ is monotonically increasing, and we are left with
a unique phase of black holes. The behaviour near this critical point is
completely analogous to a liquid/vapour system described by the
Van~der~Waals equation.\\
In the Kerr-Newman-AdS case, one can repeat this analysis along the same
lines; for $0<Q<Q_c$, where $Q_c\approx0.166$, we
obtain a behaviour analogous to that of the Kerr-AdS black hole, with the
difference that the stable small black hole branch appears already at zero
angular momentum \cite{Chamblin:1999tk}. It is then possible to obtain the
critical value $J_c(Q)$, where the point of inflection
is located, as a function of the
charge. For $Q=Q_c$ the small black hole phase disappears, and for larger
values of the charge the function $T(S)$ is monotonically increasing for
any angular momentum, and hence a unique stable black hole phase exists;
the curve of second order critical points ends in $J=0$, $Q=Q_c$.
\begin{figure}[htb]
\centerline{
\epsfig{file=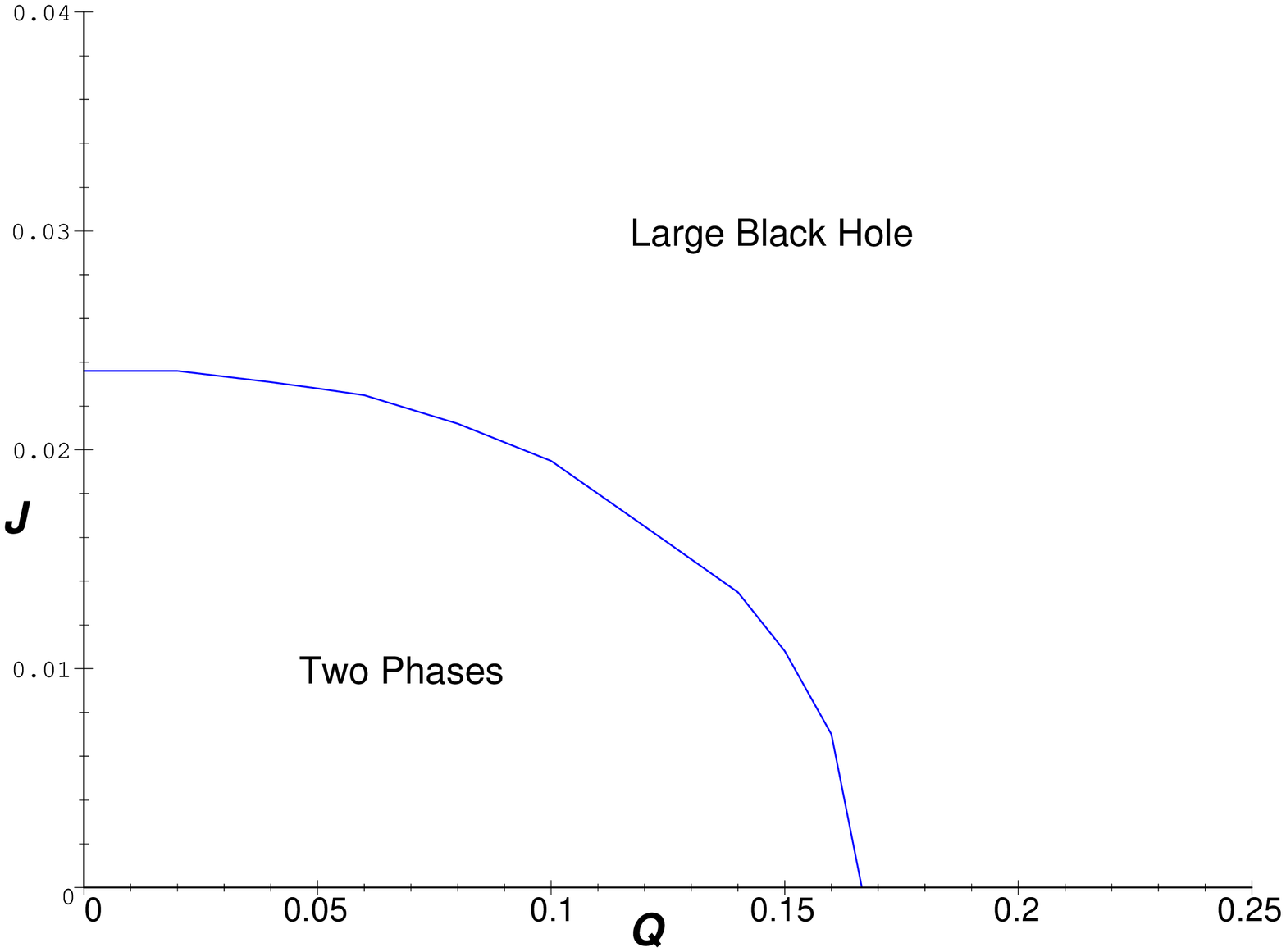,angle=-90,width=0.5\linewidth}
\epsfig{file=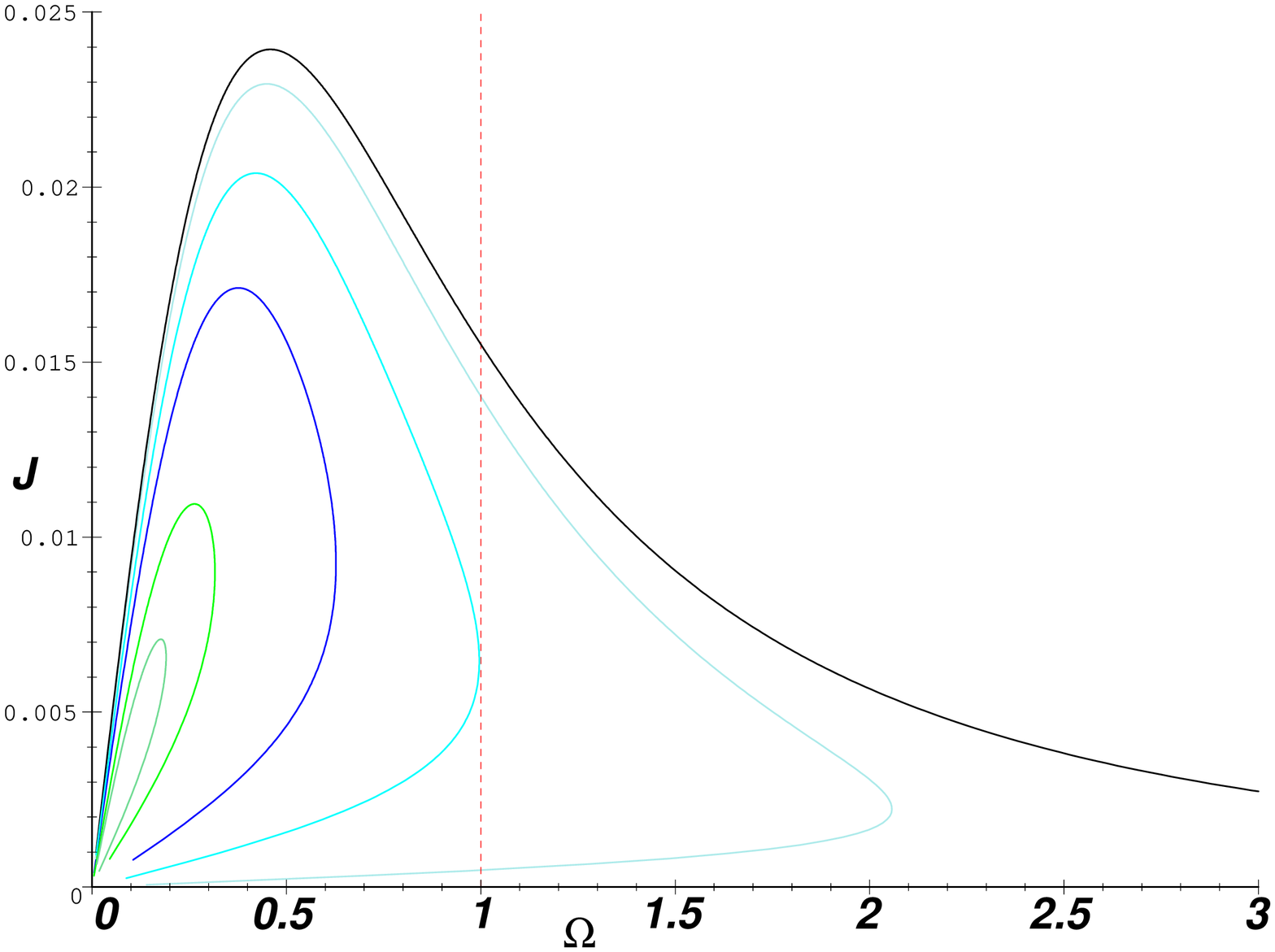,angle=-90,width=0.5\linewidth}}
\caption{{\sf Phase Diagram in the $(Q,J)$-plane and stability diagram in
    the $(\Omega,J)$-plane.} In the left plot the curve corresponds to a
    line of second order critical points. Outside the region bounded by it,
    there is a unique black hole phase. Inside the region, one has
    a small black
    hole phase at low temperatures and a large black hole phase at high
    temperatures, separated by a first order phase transition.
    In the right plot we have drawn the lines where the thermal capacity
    diverges, for $Q=0$, $0.05$, $0.091$, $0.12$, $0.15$ and $0.16$
    respectively from the outer to the inner curve. Outside these regions the
    thermal capacity is positive. The vertical line represents
    the limit where $\Omega=1/l$; for $Q>0.091$ the entire
    instability region has $\Omega<1/l$.}
\label{fig:phase}
\end{figure}
The resulting phase diagram is shown in figure \ref{fig:phase}; the solid
line corresponds to the second order critical points. Inside the curve we
have a region where two phases of black holes are allowed, with a first
order transition between a small black hole phase and a large black
hole phase as the temperature is increased, the two phases being separated
by a coexistence curve.\\
In the second graph of figure \ref{fig:phase} we have visualized the
instability region in the $(\Omega,J)$-plane. In the uncharged case we have
stable black holes for large $J$ and thermodynamic instability below the
curve. As $Q$ grows, the instability region shrinks and disappears for the
critical charge $Q_c$. The points inside the instability region correspond
to mixed phases of small/large black holes, obtained by means of a
Maxwell-like construction.\\
\begin{figure}[htb]
\centerline{
\epsfig{file=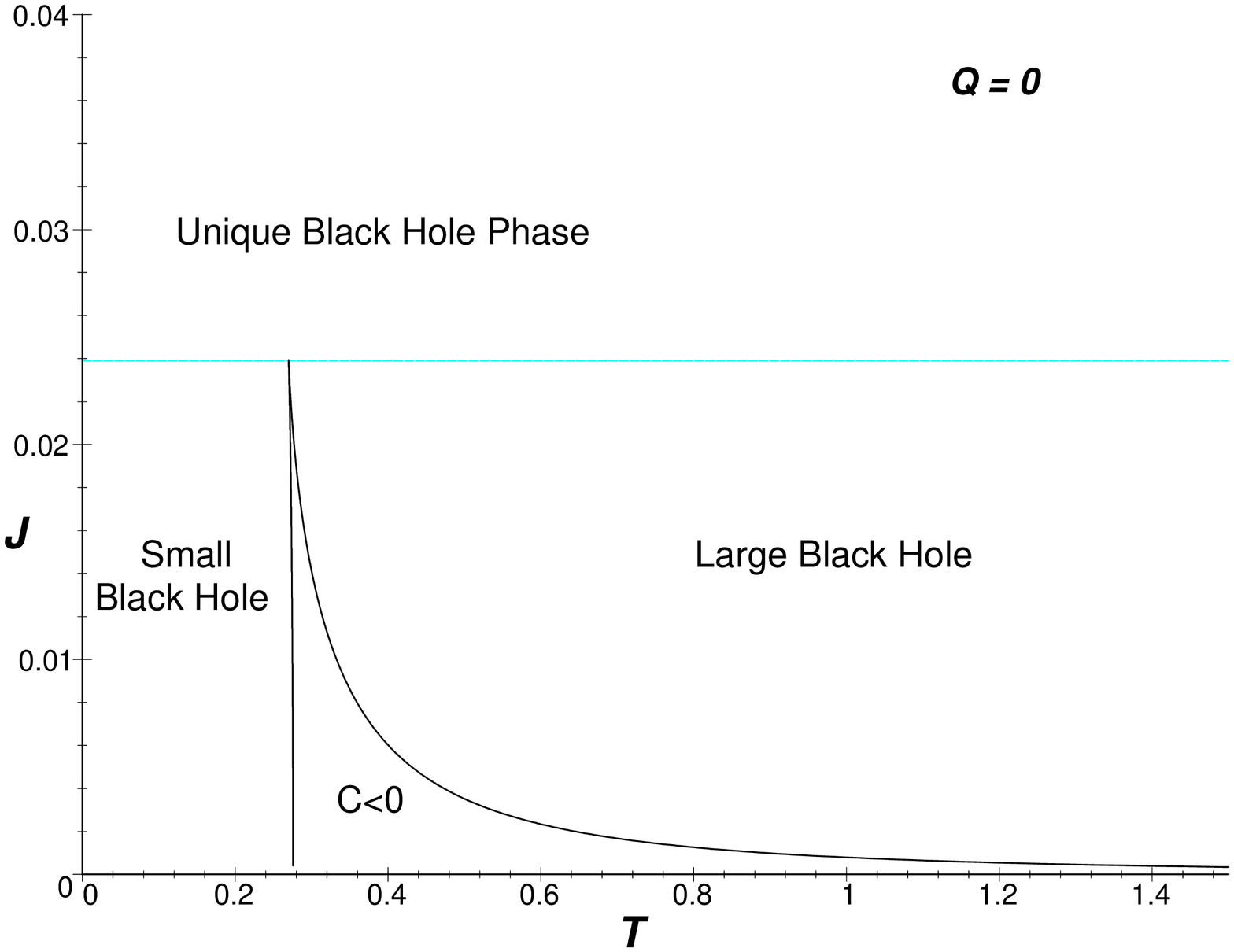,angle=-90,width=0.5\linewidth}
\epsfig{file=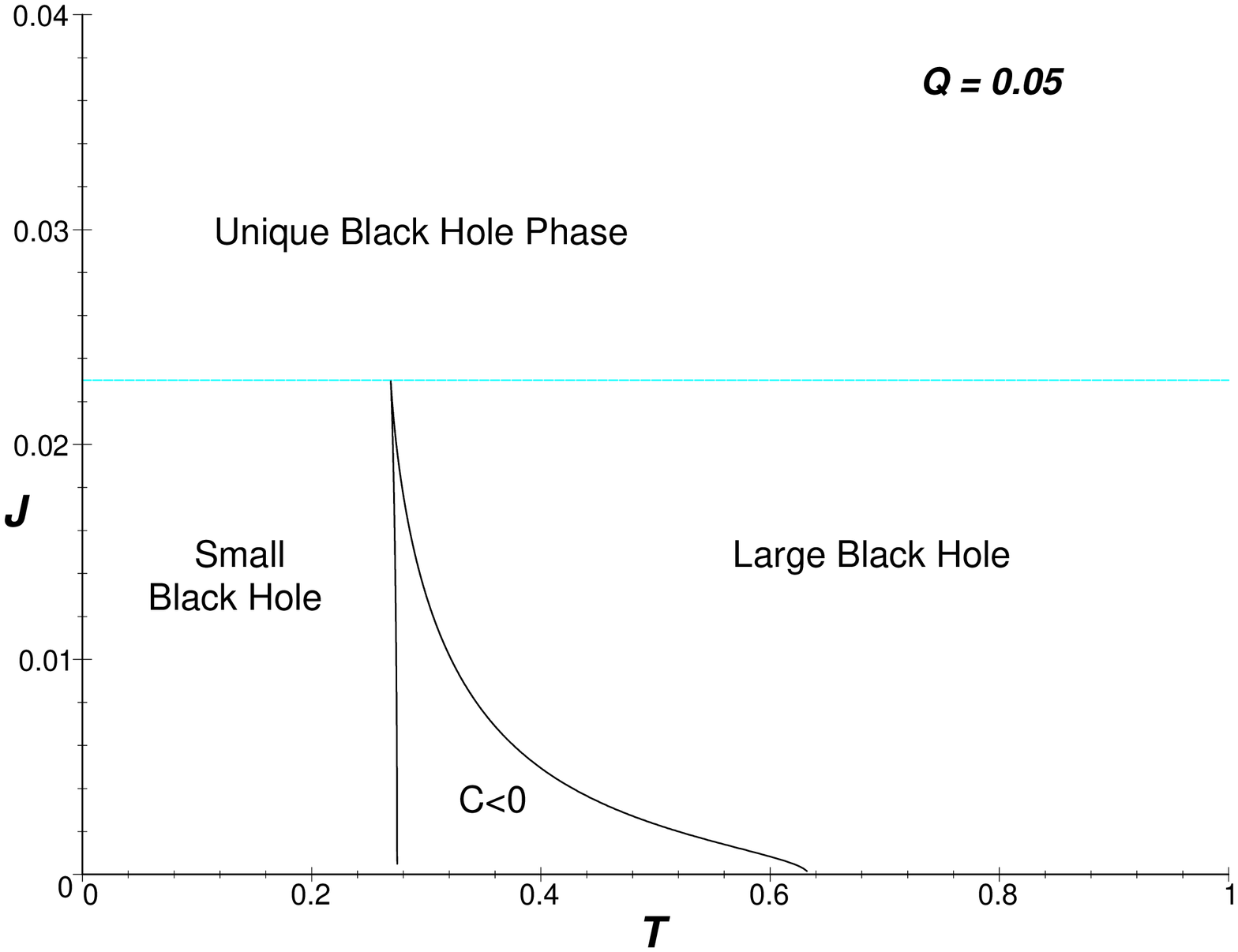,angle=-90,width=0.5\linewidth}}
\caption{{\sf Phase diagrams in the $(T,J)-$ plane.} The phase diagram of
  the $Q=0$ Kerr-AdS black hole is plotted on the left, while the diagram for
  Kerr-Newman-AdS black holes with charge $0<Q<Q_c\approx0.166$ is shown on
  the right. For higher charge the diagram is trivial, as one is left
  with a unique stable black hole phase. The left branch of the solid curve 
  is the first order small/large phase transition curve. It ends in a cusp,
  where a second order phase transition occurs.} 
\label{fig:TJ}
\end{figure}
The phase diagrams in the $(T,J)$-plane for the canonical
ensemble are shown in figure
\ref{fig:TJ}. The left diagram corresponds to the uncharged Kerr-AdS black
hole, while the one on the right side is plotted for $Q=0.05$ and shows the
qualitative behaviour of the Kerr-Newman-AdS black holes with electric
charge ranging in the interval $0<Q<Q_c$.
In these diagrams, we have plotted the curves on which the specific heat
diverges. Outside the region bounded by them, every point $(T,J,Q)$ of
the diagram corresponds to a single black hole solution, which is locally
stable (its thermal capacity is positive). In contrast, inside
the regions bounded
by these curves, which are indicated by $C<0$ in the figure,
every point $(T,J,Q)$
corresponds in fact to three different black hole solutions. To elucidate
this point let us go back to the curves $T(S)$ of figure \ref{fig:ST}. The
divergence of the thermal capacity obviously corresponds to the local
extrema of $T(S)$. For every temperature between that of the local minimum and
that of the local maximum of $T(S)$, we have three black holes: a small
metastable black hole, an unstable intermediate one, and a large stable one,
which are represented by a single point in the $(T,J,Q)$-diagram. If we
start from a small stable black hole and increase its temperature while
holding
$J$ and $Q$ fixed, as we reach the temperature of the local
minimum of $T(S)$, it
undergoes a first order phase transition with a jump in the entropy, and
reaches the black hole represented by the local minimum of $T(S)$. Then, the
state will follow the branch of large stable black holes. Hence the low
temperature branch of the $C_{JQ}=\infty$ curve corresponds to a first
order phase transition line between a small black hole phase and a large
black hole phase. This line ends in a critical point located in the cusp,
where the transition degenerates to a second order phase transition. Then,
inside the region bounded by the $C_{JQ}=\infty$ curve, the thermodynamically
stable state is simply a locally stable large black hole phase.
Finally, for Kerr-Newman-AdS black holes with charge $Q>Q_c$ the
instability region and the small/large transition disappear, leaving a
unique phase of stable black holes.\\
The local stability is however not sufficient to ensure global stability;
the full problem is very hard to tackle, but we could at least try to
compare the
free energy of the black hole with that of some reference
background like thermal AdS space, which can lead
to a \HP phase transition \cite{Hawking:1983dh}.
In doing so, however, one encounters the problem that
pure AdS space is not a solution of the Einstein-Maxwell
equations with fixed electric charge $Q$ as boundary condition.
Furthermore, AdS space has no angular momentum, as $J=0$ for
$m=0$ (cf.~\rif{MJ}). Therefore, it is not a suitable reference
background in the canonical ensemble, which has fixed $Q$ and $J$.
One should
compare the free energy of the black hole with that of other possible
solutions, such as AdS space filled with a gas of charged particles
carrying also angular momentum $J$,
or a
Kerr-Newman-AdS black hole with part of the total charge and angular
momentum carried by a gas
of such particles gravitating outside its horizon \cite{Chamblin:1999hg}.
Only for $Q=J=0$, i.~e.~the Schwarz\-schild-AdS black hole,
pure AdS space
contributes to the path integral and should be taken into account, as
it was done in \cite{Hawking:1983dh}. In the charged rotating case,
however, the question of global stability and \HP phase transitions
remains unsettled.\\
Finally, we have to examine the stationarity condition $\Omega<1/l$, which
is needed to have thermodynamic equilibrium, and a well-defined holographic
thermal state in the boundary CFT. The curve $\Omega=1/l$ can be defined
parametrically in the $(T,J)$-plane by the equations
\begin{eqnarray}
J^2&=&\frac S{4\pi^4(S+\pi)}\lp S^2+\pi S+\pi^2 Q^2\rp^2\:,\nonumber\\
T^2&=&\frac1{16}\frac{(2S+\pi)^2\lp S^2+\pi S-\pi^2Q^2\rp^2}
{\pi^2S^3(S+\pi)^3}\:,
\end{eqnarray}
with 
\eq
S\in\left[\frac\pi2\lp\sqrt{1+4Q^2}-1\rp,+\infty\right[\:.
\feq
The lower bound corresponds to supersymmetric black holes for $Q\neq0$, and for
$S\rightarrow\infty$ the curve is always asymptotic to $T=1/(2\pi l)$. 
\begin{figure}[htb]
\centerline{
\epsfig{file=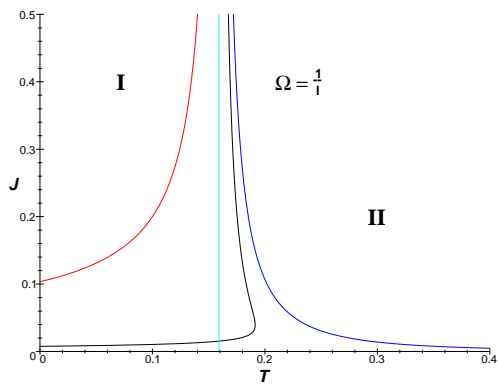,angle=-90,width=0.5\linewidth}}
\caption{{\sf $\Omega=1/l$ curves.} These curves have been plotted for the
  values $Q=0$, $Q=0.2$ and $Q=0.5$ of the charge, from the right to the left
  respectively. The vertical line is the asymptote $T=1/(2\pi l)$. For charges
  $0<Q<1/\sqrt8$ the qualitative behaviour is similar to the intermediate
  curve, while for higher charges the corresponding curves behave like the
  one with $Q=0.5$. In region $I$ we have $\Omega>1/l$ and in region $II$ we
  have $\Omega<1/l$. The intersection of these curves with the $T=0$ axis
  corresponds to supersymmetric black holes.} 
\label{fig:Omega}
\end{figure}
The behaviour of the curve depends on the electric charge, and has been
summarized in figure \ref{fig:Omega}.
For $Q=0$, the curve is strictly decreasing from the asymptote and ends at
$J=0$ for infinite temperature.
If $0<Q<1/\sqrt8$, the infinite temperature branch is deformed, curves
back towards the $J$ axis and intersects it in
the supersymmetric point.
Finally, for $Q>1/\sqrt8$, the curve remains at $T<1/{2\pi l}$, and is strictly
increasing from its intersection with the ordinate axis, again describing a
supersymmetric black hole, to the asymptote.
In any case the $\Omega=1/l$ curve splits the $(T,J)$-plane into two
regions, named $I$ and $II$ in figure \ref{fig:Omega}. Black holes in 
region $I$, at the left of the $\Omega=1/l$ curve, spin at angular
velocity $\Omega$ greater than $1/l$ and show superradiant instability.
On the other hand, black holes in region $II$, at the right of the
curve, have $\Omega<1/l$, and have a well-defined thermodynamics.
However, this is not the whole story. Inside the regions of local
instability shown in figure \ref{fig:TJ}, every point $(T,J,Q)$ corresponds
in fact to three black hole solutions: an unstable, a metastable and a
locally stable black hole. It is possible to verify that the locally stable
one spins at $\Omega<1/l$, and hence the whole region of $\Omega<1/l$ black
holes is the union of region $II$ and the instability region. This stems
from the fact that the locally stable black hole, at fixed $(T,J,Q)$, is
the one with maximum entropy, and that $\Omega(S,J,Q)$ is a decreasing
function of $S$ at fixed $J$ and $Q$.
\begin{figure}[htb]
\centerline{
\epsfig{file=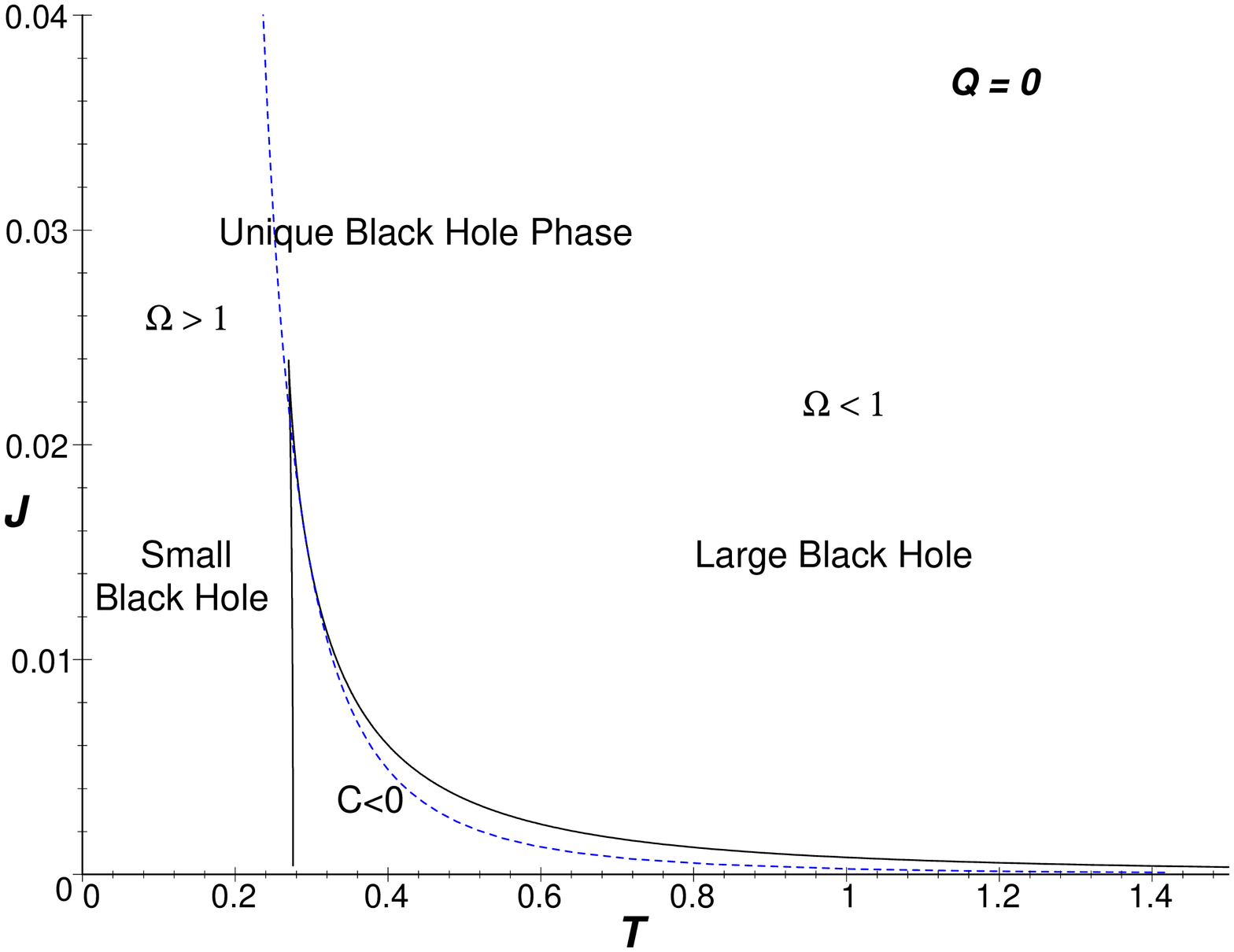,angle=-90,width=0.5\linewidth}
\epsfig{file=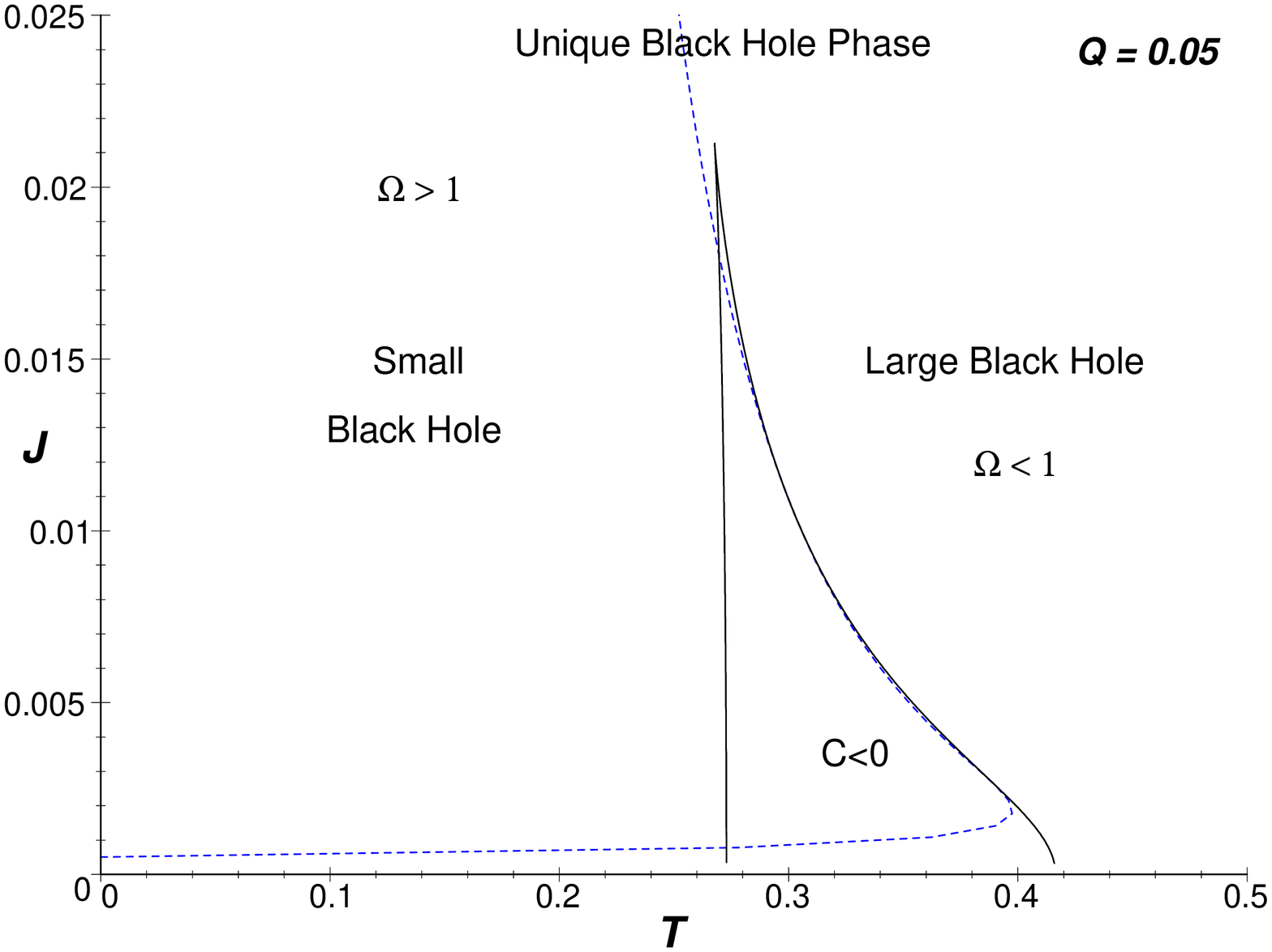,angle=-90,width=0.5\linewidth}}
\caption{{\sf Phase diagrams with $\Omega=1/l$ curve,
  for charges $Q=0$ and $Q=0.05$.}
  The second diagram is characteristic for all black
  holes with $0<Q<Q_c$. The dotted curve represents the line $\Omega=1/l$.}
\label{fig:phase2}
\end{figure}
\\The full phase diagram, taking into account the restriction on the angular
velocity, is reported in figure \ref{fig:phase2}. It implies that part of
the small/large black hole phase transition has no dual CFT analogue, as in
this region one has $\Omega>1/l$, and so the Einstein universe in
which the dual CFT lives would have to rotate faster than light.
However the second order critical point lies
in the $\Omega<1/l$ region. Furthermore, for charged black holes, the first
order phase transition occurs at $\Omega<1/l$ for sufficiently small
angular momentum, and has therefore an analogue on the CFT side.

\subsection{Grand-Canonical Ensemble}

As described in Section \ref{potentials}, the thermodynamics of the
grand-canonical ensemble can be extracted from the Gibbs potential
$G(T,\Omega,\Phi)$. The natural variables for this ensemble are the
intensive parameters $T,\Omega$ and $\Phi$; however, to simplify the
analysis, we shall study the thermodynamics in the $(r_+,a)$-plane, at
fixed electric potential $\Phi$.\\
In this plane, the black holes with genuine event horizon are restricted to
the region $a_{\mathrm extr}^-(r_+,\Phi)\leq a\leq
a_{\mathrm extr}^+(r_+,\Phi)$ with $a<l$,
where
\eq
a_{\mathrm extr}^\pm(r_+,\Phi)=\frac{r_+}{2l^2|\Phi|}\left(
r_+^2-l^2+2l^2\Phi^2\pm\sqrt{(r^2_+-l^2)^2+16r_+^2l^2\Phi^2}
\right)^{1/2}.
\feq
Between these two curves the black holes have a bifurcate horizon, on the
curves they are extremal, and outside the solution shows a naked
singularity. If $|\Phi|\leq1$, $a^-_{\mathrm extr}$ is negative, and the entire
region below the curve $a_{\mathrm extr}^+(r_+,\Phi)$ represents black hole
solutions. For $\Phi>1$, $a^-_{\mathrm extr}$ is positive for 
$0\leq r_+\leq l\sqrt{(\Phi^2-1)/3}$; if $r_+$ lies in this
range, $a_{\mathrm extr}^-$ determines a new region
with naked singularities, and the black
holes are found between the two curves. For $r_+$ outside of this interval,
all the points represent black holes, as long as $a<l$. Finally, in the limit
$\Phi\rightarrow\infty$, the two curves merge and only a line of extremal
black holes with $r_+=a<l$ survives.\\
Another relevant information in the thermodynamic diagram is given by the
region where the Killing vector $\partial_t + \Omega_H\partial_{\phi}$
is timelike all the way out to infinity.
It is easy to see that this
condition is satisfied for $a\leq r_+^2/l$, and so is always fulfilled for
$r_+>l$, as we assumed $a<l$.
One readily verifies that the condition $a\leq r_+^2/l$ is equivalent
to $\Omega < 1/l$.
In the diagrams, the region at the right of the curve $a=r_+^2/l$
is the region
where the black hole can be in thermal and dynamical equilibrium with
radiation, and the rotating Einstein universe at the boundary spins slower
than light.\\
Let us now turn to the stability analysis of the black hole. We
shall begin with local stability. The curve separating stable
from unstable regions is given by Eq. \rif{qe2}. In fact this equation can
be solved for $a(r_+,\Phi)$ after having substituted $q$
in terms of $\Phi$ by means of \rif{Phi1}; the result is quite complicated,
but nevertheless adequate for visualizing the local stability regions in the 
$(r_+,\Phi)$-plane.\\
However locally stable black holes can decay into other states with lower
free energy. In the case of the grand-canonical ensemble, another possible
state is given by AdS space, which can have arbitrary electric potential
$\Phi$, constant all over the spacetime. The Gibbs free energy of AdS space
filled with thermal radiation vanishes identically, hence the global
stability can be investigated by studying the sign of $G(T,\Omega,\Phi)$.
The free energy of the black hole is negative for
\eq
r^2_+(a,\Phi)\geq \frac{l^2}2\left(1-\Phi^2+\sqrt{\left(1-\Phi^2\right)^2+\frac{4a^2}{l^2}\Phi^2}\right)\equiv r^2_{\mathrm HP}(a,\Phi),
\feq
which means that the black hole dominates in this region.
For $r_+=r_{\mathrm HP}(a,\Phi)$ there is a first
order phase transition line, with a discontinuity in the entropy, and for 
$r_+<r_{\mathrm HP}(a,\Phi)$ the AdS solution is globally preferred. However we
stress that this does not exhaust the question of global stability for
these black holes, as other solutions with lower free energy may
exist \cite{Chamblin:1999hg}. Indeed, there are regions in the
$(r_+,a)$-plane in which the black hole, though being
unstable, has lower free
energy that AdS space; it is plausible that a new solution exists which
minimizes the Gibbs potential.\\
\begin{figure}[htb]
\centerline{
\epsfig{file=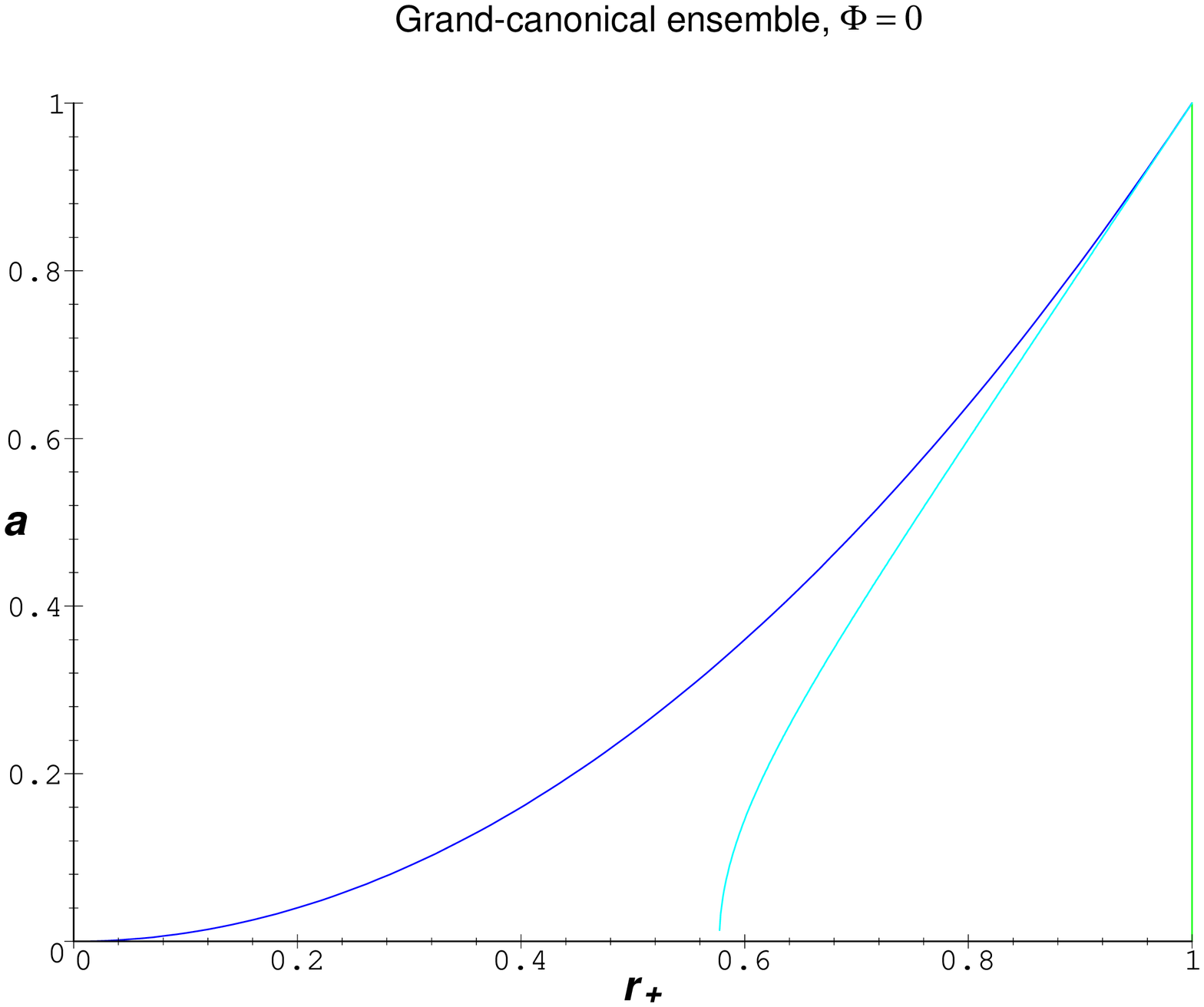,angle=-90,width=0.5\linewidth}
\epsfig{file=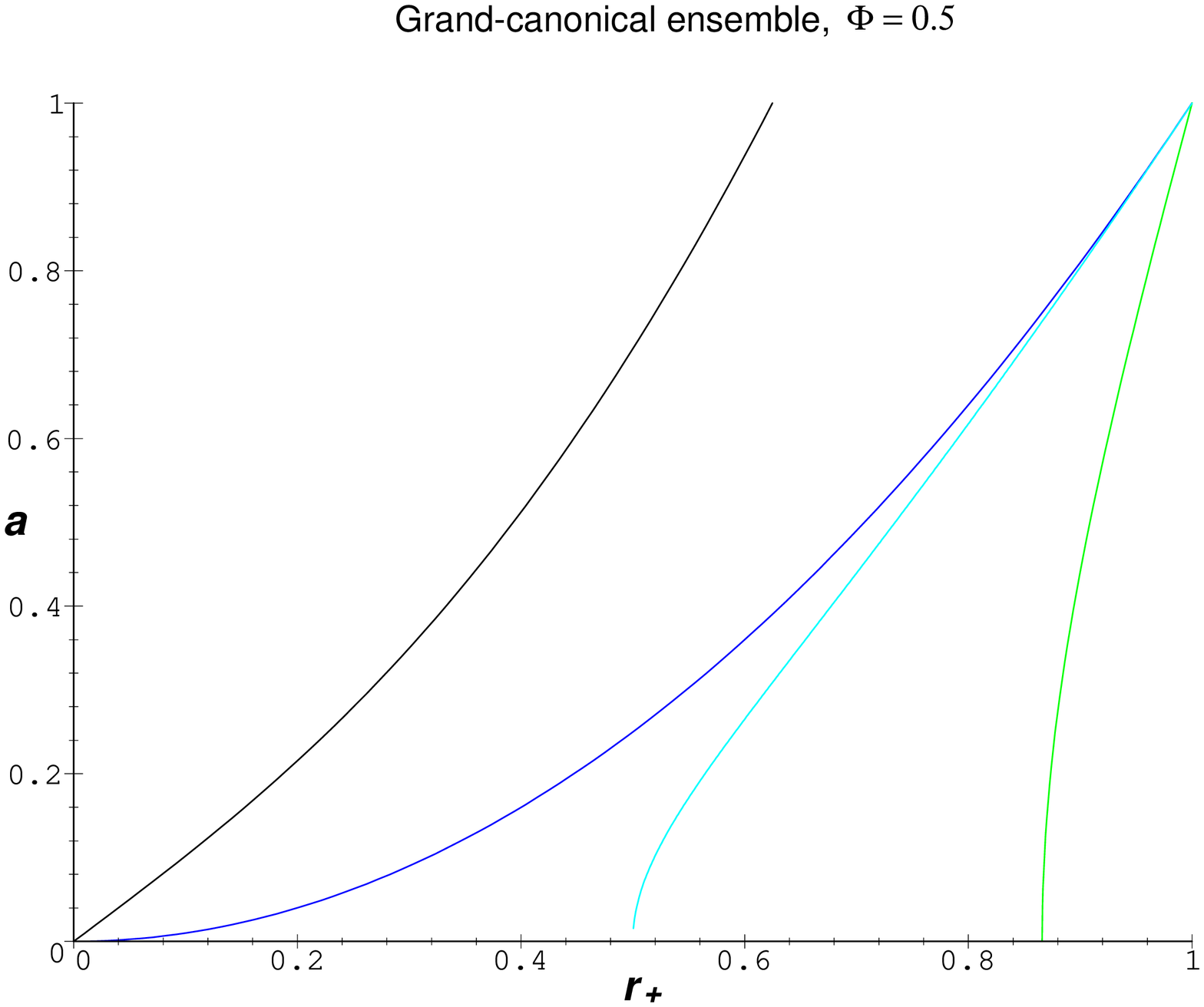,angle=-90,width=0.5\linewidth}}
\centerline{
\epsfig{file=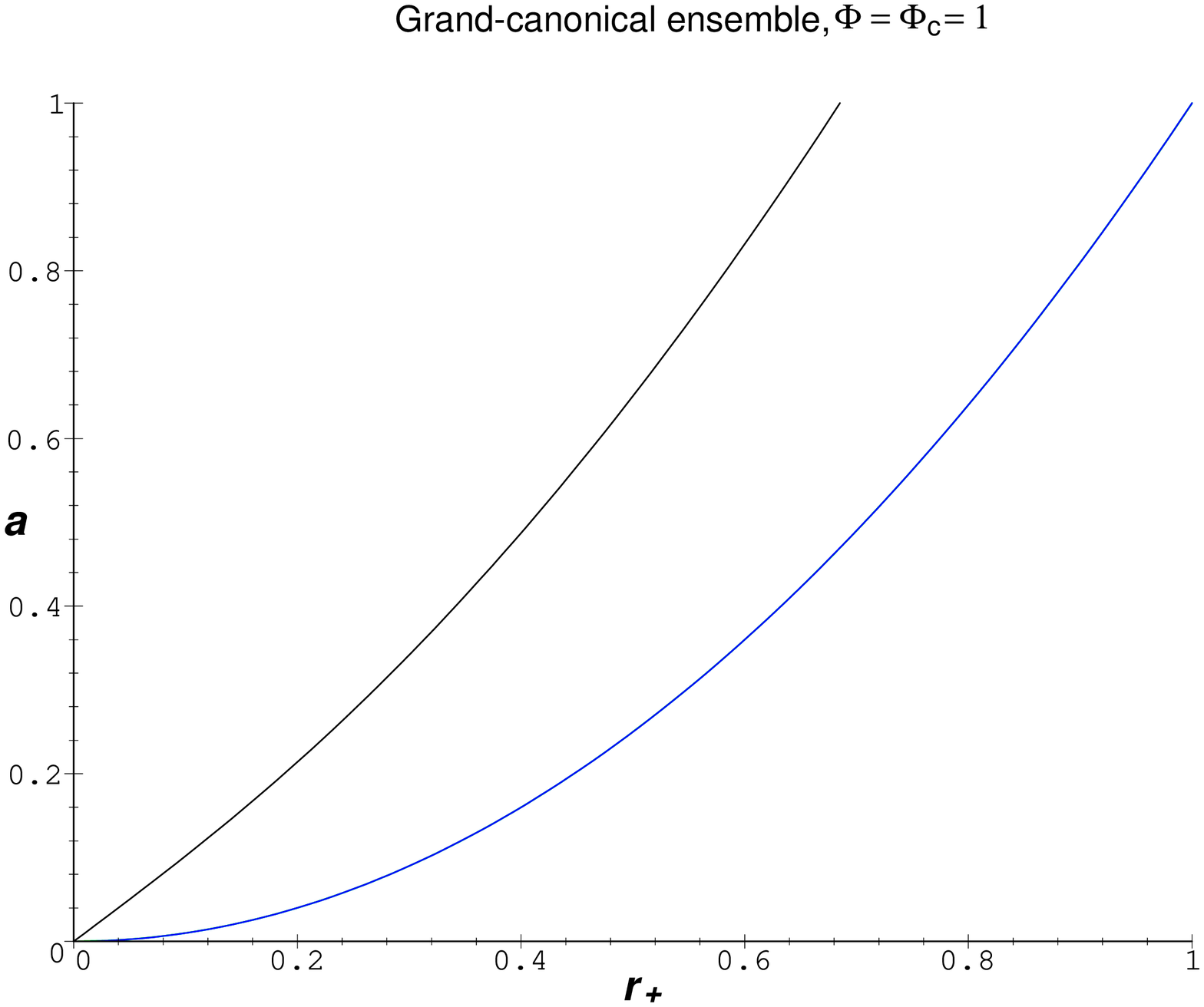,angle=-90,width=0.5\linewidth}
\epsfig{file=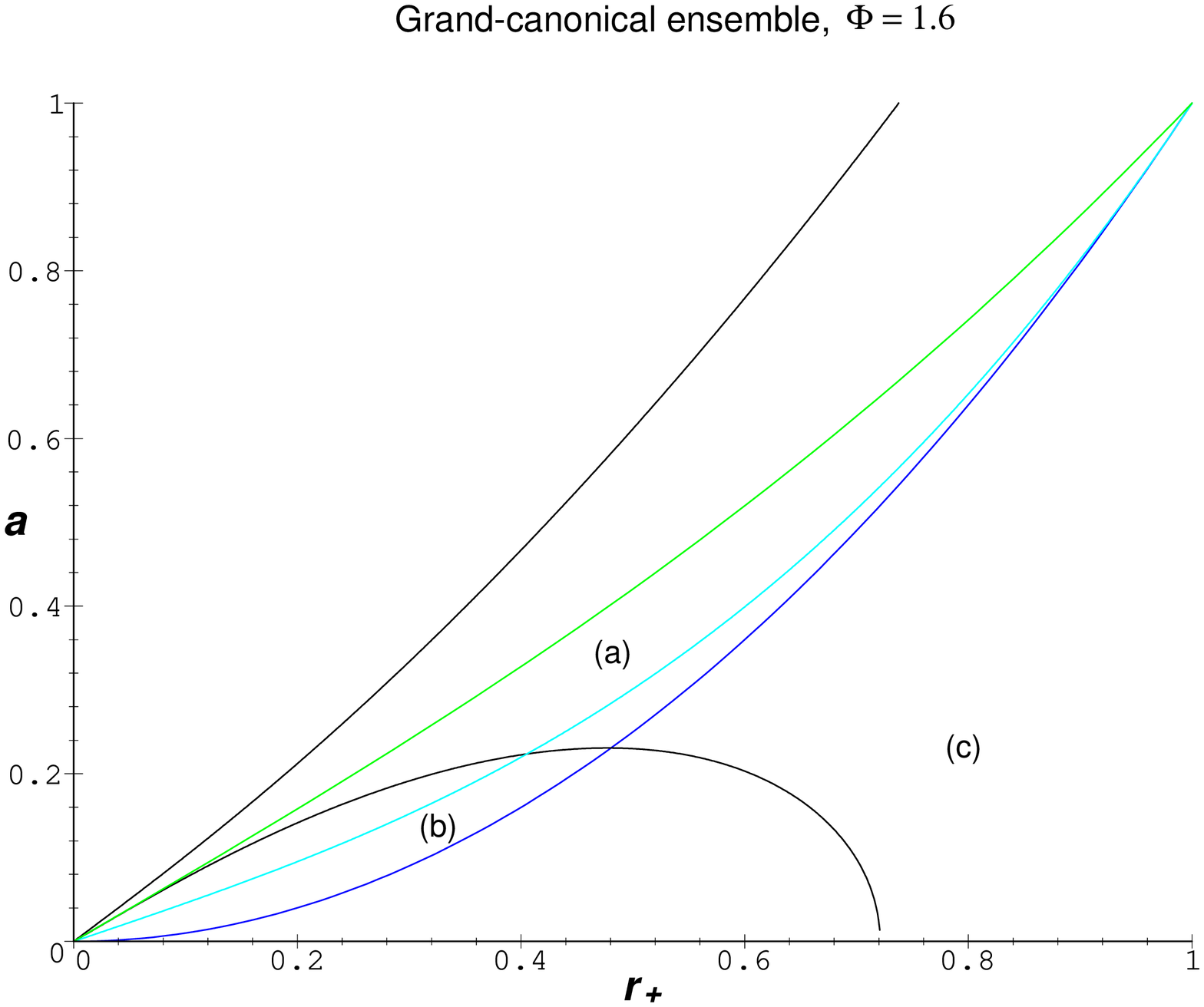,angle=-90,width=0.5\linewidth}}
\caption{{\sf Phase structure in the grand-canonical ensemble in the
  $(r_+,a)$-plane, for various values of the electric potential $\Phi$.}
  For $\Phi=0$ we have, from the left to the right, the $\Omega=1/l$
  curve, the line bounding the stability region, and the Hawking-Page
  transition curve. For $0<\Phi<\Phi_c=1$, in addition to these three lines,
  the extremality curve appears at the left. For the critical potential
  $\Phi=\Phi_c$, the $\Omega=1/l$ curve, the stability bound, and the
  Hawking-Page transition line coalesce. Finally, for $\Phi>\Phi_c$,
  an additional extremality line appears (the one intersecting the
  $r_+$ axis twice), and the order of the $\Omega=1/l$ curve, the
  stability bound, and the Hawking-Page transition line is reversed.
  The various regions bounded by these curves are discussed in the text.}
\label{fig:gc}
\end{figure}
Finally we observe that the $\Omega=1/l$ curve, the local stability curve
and the Hawking-Page transition curve intersect at the point $(r_+=l,a=l)$
for any choice of the electric potential. For $\Phi=\Phi_c=1$
these three curves
coincide; for $\Phi<\Phi_c$ we find the $\Omega=1/l$ curve,
the local stability curve and the Hawking-Page transition curve
in this order from the left to the right,
while for $\Phi>\Phi_c$ they appear in the reverse order.\\
We have summarized this analysis in figure \ref{fig:gc}, which shows
the phase structure in the $(r_+,a)$-plane
for various values of the
electric potential.
We see that for $\Phi<\Phi_c=1$, the Hawking-Page transition curve is the
rightmost one; it separates a large $r_+$ phase of
large black holes and an AdS phase, which dominates over
black holes with small $r_+$.
Inside the AdS phase region, with increasing $r_+$, we first have a zone
of naked singularities separated by the extremality line from a region with
genuine black holes, which however do not possess a boundary CFT
analogue,
as they have $\Omega>1/l$, and which are furthermore
locally unstable. In the next region we find locally unstable black holes
which rotate slower than light, and finally a region with locally stable
but globally unstable black holes which ends on the Hawking-Page transition
curve. These regions are anyway dominated by the AdS phase.\\
For $\Phi>\Phi_c$ the situation is more subtle; a new branch of extremal
black holes appears and bounds a region of naked singularities for small
$a$. For small $r_+$ we again have the AdS dominated phase, but
at the right of the Hawking-Page transition curve
we find (a) a region with locally
unstable black holes with $\Omega > 1/l$, having however negative
Gibbs free energy, so they are preferred with respect to AdS space,
(b) a region with
stable black holes with $\Omega>1/l$, and finally (c) a region of stable black
holes with $\Omega<1/l$. This raises the question for region (a) whether
there exists at all a thermodynamically stable state for this range of
parameters. Furthermore, the region defined by
$a<a_{\mathrm extr}^-(r_+,\Phi)$ represents naked singularities
which have a lower Gibbs free energy than AdS space itself.

\section{Concluding Remarks}
\label{concl}

In the present paper, we studied the thermodynamics of the
four-dimensional Kerr-Newman-AdS black hole both in the canonical
and the grand-canonical ensemble. Thereby we encountered an
interesting phase structure. In the canonical ensemble, we found
a small-large black hole first order phase transition, which 
disappears for sufficiently large electric charge or angular
momentum. In the points where the small-large transition disappears,
the first order phase transition degenerates to a higher order one.
As pure AdS space is not a solution of the Einstein-Maxwell field
equations with fixed electric charge, we cannot compare the black hole
Helmholtz free energy with that of pure AdS space, in order to
study global stability. It would therefore be desirable to have
another reference background to our disposal, which describes AdS
space filled with charged rotating particles.
Another possibility
would be to have a KNAdS black hole surrounded by a gas of particles
carrying part of the total charge and angular momentum
(cf.~ the discussion in \cite{Chamblin:1999hg}).

As AdS space can have arbitrary constant electric potential,
we can compare the Gibbs free energy of the black hole with
that of AdS space in the grand-canonical ensemble. In this way, we
found the Hawking-Page transition curve. Also in this context,
some intriguing features arise, e.~g.~one finds black holes which
are locally unstable, but nevertheless have lower Gibbs free energy than
AdS space itself. There is also a region of naked singularities
in the $(r_+,a)$-phase diagram, which are energetically preferred
with respect to AdS space. It would be interesting to study if these
naked singularities play some role in the dual CFT.

We hope to have clarified that a consistent treatment of rotating
AdS black hole thermodynamics requires the usage of the angular velocity
$\Omega$ of the rotating Einstein universe, in which the dual conformal field
theory lives. This angular velocity differs from that of the event horizon,
which enters the thermodynamical description of asymptotically flat
black holes. The reason for this discrepancy is that the angular velocity
\rif{om} appearing in the canonical form of the metric, does not
vanish at infinity. (It even has the opposite sign of $\Omega_H$,
cf.~\rif{om}). The fact that a consistent thermodynamics can be
formulated using $\Om$ is in full agreement with the AdS/CFT correspondence.

The black holes we have considered can also be lifted to eleven dimensions,
as solutions of ${\cal N}=1$, $D=11$ supergravity
\cite{Chamblin:1999tk,Cvetic:1999xp}. 
These solutions represent the
near-horizon limit of M2-branes with spherical
worldvolumes, rotating both about an axis of the worldvolume and in planes
orthogonal to the worldvolume, the latter rotation
giving rise to four R-charges upon Kaluza-Klein compactification. The
Kerr-Newman black holes considered here correspond to the special case
where all transverse angular momenta of the M2-brane, and hence all
R-charges of the black hole, are equal
\footnote{A thermodynamical
discussion of branes rotating in planes orthogonal
to the worldvolume can be found in
\cite{Gubser:1998jb,Kraus:1998hv,Cai:1998ji,Cai:1999hg,Hawking:1999ab}.
In \cite{Gubser:1998jb}, a field theory model of rotating D3-branes
was proposed. This was later generalized in \cite{Cvetic:1999rb}
to rotating M-branes.}.

The thermodynamic structure we have elaborated, corresponds then to
that of the worldvolume field theory of $N$ coincident such branes.

It would be very interesting to give a microscopic description of
the Kerr-Newman-AdS black holes in terms of this dual CFT. Let us
recall in this context that only black holes whose angular velocity
satisfies
\eq
\Omega \le \frac 1l \label{angvelcond}
\feq
correspond to thermal states in the dual field theory, as otherwise
the rotating Einstein universe, in which the dual CFT lives, would
rotate faster than light. One easily shows that for zero temperature
black holes, condition \rif{angvelcond} implies
\eq
q^2 \ge a(a+1)^2 \label{bound}
\feq
for the electric charge and rotation parameter appearing in the metric. If the
bound \rif{bound} is saturated, we have a BPS state. It would be
interesting to study if the corresponding field theory at weak
coupling reproduces the Bekenstein-Hawking entropy of the
black hole in this BPS limit. As the M2-brane worldvolume
field theory is the strong coupling limit (conformal point)
of the D2-brane worldvolume theory \cite{Maldacena:1997re},
the relevant gauge theory
would be that on the worldvolume of D2-branes.
In any case, in the presence of a global background $U(1)$ current
(corresponding to the black hole charge $Q$),
the zero temperature state of the dual field theory must be highly
degenerate, as the extremal black hole has nonvanishing
entropy\footnote{Note that for zero $U(1)$ current, the extremal black
holes have $\Om > 1/l$ \cite{Hawking:1998kw}, and therefore do not
correspond to states in the CFT.}. It would also be interesting to
consider the five-dimensional case, corresponding to ${\cal N}=4$, $D=4$
SYM on a rotating Einstein universe \cite{Hawking:1998kw}. However,
the generalization of the rotating five-dimensional AdS black holes
found in \cite{Hawking:1998kw} to nonzero charge
seems to be a nontrivial task. If such
a solution were known, one could study if there exist BPS states
(which can only occur for nonzero charge \cite{Hawking:1998kw}), and
compare their entropy with the one obtained on the CFT side. Apart from
the possiblity of identifying black hole microstates, this would also
be a check of the AdS/CFT correspondence.




\section*{Acknowledgements}

D.~K.~has been partially supported by a research grant
within the scope of the {\em Common Special Academic Program III} of the
Federal Republic of Germany and its Federal States, mediated 
by the DAAD.\\
The authors would like to thank M.~M.~Taylor-Robinson and
L.~Vanzo for helpful discussions.


\providecommand{\href}[2]{#2}\begingroup\raggedright\endgroup
\end{document}